\documentclass[aps,prd,nofootinbib,twocolumn,showpacs]{revtex4-1}

\usepackage{multirow}
\usepackage{amsmath,epsfig,ulem}
\usepackage{amsfonts}
\usepackage{amsmath}
\usepackage{amssymb}
\usepackage{graphicx}
\usepackage{appendix}
\usepackage{hhline,color}
\usepackage{pifont}
\usepackage{lmodern}
\usepackage[utf8]{inputenc}

\newcommand{\pr}[1]{\ensuremath{\left[#1\right]}}
\newcommand{\pc}[1]{\ensuremath{\left(#1\right)}}

\newcommand{\ev}[1]{\ensuremath{\left\langle #1\right\rangle}}
\newcommand{\Z}{\mathbb{Z}}

\begin{document}

\title{Influence of the vector interaction and an external magnetic field on the 
isentropes near the chiral critical end point}

\author{Pedro Costa}
\email{pcosta@teor.fis.uc.pt}
\affiliation{Centro de F\'{\i}sica da Universidade de Coimbra (CFisUC), Department of Physics,
University of Coimbra, P-3004-516  Coimbra, Portugal}

\date{\today}

\begin{abstract}
The location of the critical end point (CEP) and the isentropic trajectories 
in the QCD phase diagram are investigated. We use the (2+1) Nambu$-$Jona-Lasinio 
model with the Polyakov loop coupling for different scenarios, namely by 
imposing zero strange quark density, which is the case in the ultra relativistic 
heavy-ion collisions, and $\beta$-equilibrium.
The influence of strong magnetic fields and of the vector interaction on the 
isentropic trajectories around the CEP is discussed.
It is shown that the vector interaction and the magnetic field, having opposite 
effects on the first-order transition, affect the isentropic trajectories 
differently: as the vector interaction increases, the first-order transition 
becomes weaker and the isentropes become smoother; 
when a strong magnetic field is considered, the first-order transition is 
strengthened and the isentropes are pushed to higher temperatures. 
No focusing of isentropes in region towards the CEP is seen.

\end{abstract}


\maketitle

\section{Introduction}

The main goal of the heavy-ion collision (HIC) program is to understand strong 
force and extended systems governed by Quantum Chromodynamics (QCD). 
Developments over the last two decades allowed the creation and investigation of 
new forms of QCD matter characterized by high parton densities. One major 
achievement in HIC was the discovery that QCD matter at energy densities greater 
than 1 GeV/fm$^3$ acts like a strongly interacting plasma of quarks and gluons.   
Indeed, the fast (local) thermalization time and the good agreement of the 
data at RHIC with ideal relativistic hydrodynamics models (which admit a fluid 
evolution with zero viscosity) are evidences that the matter formed at RHIC is a 
strongly interacting plasma of quarks and gluons \cite{d'Enterria:2006su},
as confirmed by the Large Hadron Collider (LHC) data \cite{Averbeck:2015jja}.

Since in HIC the expansion of the quark-gluon plasma (QGP) is accepted to be a 
hydrodynamic expansion of an ideal fluid, it will nearly follow trajectories of 
constant entropy, the so-called isentropes. 
Due to the conservation of the baryon number, the isentropic trajectories are 
lines of constant entropy per baryon ($s/\rho_B$) in the 
($T,\mu_B$) space with zero strange quark density, $\rho_s= 0$, which contain 
important information on the adiabatic evolution of the system. 
For AGS, SPS, and RHIC the values of $s/\rho_B$ are 30, 45, and 300, respectively 
\cite{Bluhm:2007nu}.
Lattice results for the isentropic (2+1) flavors equation of state (EOS) at these 
values of $s/\rho_B$ are given in Refs. \cite{DeTar:2010xm,Borsanyi:2012cr}.

In Ref. \cite{Asakawa:2008ti} it was proposed that the presence of a critical end 
point (CEP) in the QCD phase diagram deforms the trajectories describing the 
evolution of the expanding fireball. 
This will have important consequences on the search for the CEP, because 
modifications of the expansion trajectory may lead to observable effects in 
the hadron spectra (see Ref. \cite{Senger:2011zza}).

The possible existence of the CEP and its implications to the QCD phase diagram 
is a very timely topic that has drawn the attention of the physics community. 
From the theoretical point of view, the location of the CEP has been 
intensively investigated by using lattice QCD calculations (despite the fermion 
sign problem, extrapolation methods have been used to access the region of small 
chemical potentials and look for the CEP \cite{Endrodi:2011gv}), 
and more recently by using Dyson-Schwinger equations \cite{Fischer:2014ata}.
Also, effective models such as the Nambu$-$Jona-Lasinio (NJL) model and its extensions, 
like the NJL model with eight-quark interactions and the 
Polyakov$-$Nambu$-$Jona-Lasinio (PNJL) model, have been used to study critical properties 
around the CEP \cite{Costa:2008yh,Hiller:2008nu,Costa:2008gr}.

From the experimental point of view, the location of the CEP is one major goal 
of several HIC programs, but so far, no definitive results have been found about its  
location, and even its existence remains a mystery.
Since 2010, the Beam Energy Scan (BES-I) program at RHIC has been searching for the 
experimental signatures of the first-order phase transition and the CEP by 
colliding Au ions at several energies \cite{Abelev:2009bw}. In the near future 
it is expected that, if the CEP exists at a baryonic chemical potential below 
400 MeV, the upcoming BES-II program can provide data on fluctuation and flow 
observables which should yield quantitative evidence for the presence of the CEP.
Also, at RHIC, the STAR Collaboration is looking for the CEP, but no definitive 
conclusions were possible from their measurements of the moments of 
net-charge multiplicity distributions. Future measurements with high 
statistics data will be needed \cite{Adamczyk:2014fia}.

The NA49 program at CERN SPS has also investigated the CEP's location in nuclear 
collisions at 158$A$ GeV by analyzing $\pi^+ \pi^-$ pairs with an invariant mass 
very close to the two-pion threshold, looking for critical fluctuations of the 
sigma component in a hadronic medium \cite{Anticic:2009pe}.
Sizable effects of $\pi^+\pi^-$ pair fluctuations with critical characteristics 
were found in Si + Si collisions, but these effects could not be directly related 
to the presence of the CEP.
Presently, the NA61/SHINE program is dedicated to looking for the CEP and 
investigating the properties of the onset of deconfinement in light and heavy ion 
collisions \cite{Gazdzicki:2011fx,Aduszkiewicz:2015jna}.

Recently, new possible hints on the CEP were given: 
\begin{itemize}
\item[i)] in \cite{Lacey:2014wqa} a finite-size scaling analysis of non-monotonic 
excitation functions for the Gaussian emission source radii difference obtained 
from two-pion interferometry measurements in Au+Au 
($\sqrt{s_{NN}} = 7.7-200$ GeV) and Pb+Pb ($\sqrt{s_{NN}} = 2.76$ TeV) collisions 
suggests a second order phase transition with the estimated location of the CEP 
at $T^{\,CEP} = 165$ MeV and $\mu_B^{CEP}= 95$ MeV; 
\item[ii)] in \cite{Bugaev:2014bua} it was observed a sharp peak in the trace-anomaly 
and a local minimum of the generalized specific volume at a laboratory energy of 
11.6 AGeV which can provide a signal for the formation of a mixed phase between 
the quark-gluon plasma and the hadron phase; 
\item[iii)] in \cite{Nara:2016phs} it is argued that the observed collapse of directed 
flow of protons and pions at mid-rapidity at 9 GeV $<\sqrt{s_{NN}}<20$ GeV is 
the evidence for the softening of the QCD equation of state (EoS), possibly 
caused by a first-order phase transition. 
\end{itemize}
If these results are confirmed, i.e., an unambiguous experimental identification 
of a first-order phase transition from the hadronic phase to the deconfined phase,
and having in mind that lattice QCD calculations at zero baryon chemical potential 
show an analytic crossover for the transition from hadrons to quarks and gluons 
\cite{Aoki:2006we,Aoki:2009sc,Borsanyi:2013bia}, the existence of the CEP gains a strong 
foothold.

As a matter of fact, it might be easier to detect the first-order phase-coexistence 
region than the CEP: when the expanding matter created in a HIC eventually 
crosses a first-order phase transition line the system probably will 
spend sufficient time in this region to develop measurable signals.
Possible observables of this transition are based on the clumping of the 
system due to spinodal phase decomposition. 
This could lead to enhanced fluctuations of observables like the strangeness:
the enhancement of the kaon-to-pion fluctuations can be the result of the 
enhancement of fluctuations in the strangeness sector (for a review, see Ref. 
\cite{Senger:2011zza}).
These fluctuations caused by spinodal instabilities are a generic phenomenon 
of first-order phase transitions and may be less suppressed by the short lifetime 
and the finite size of the system, as compared to critical fluctuations needed
to detect the CEP. 

The experience acquired in theoretical and experimental investigations 
concerning the nuclear liquid-gas phase transition can be very useful to the
study of deconfinement and chiral phase transitions in relativistic HIC
\cite{Mishustin:2006ka}. 
Indeed, in Ref. \cite{Borderie:2001jg}, the spinodal instabilities as a signal of the 
nuclear liquid-gas phase transition have been successfully identified in nuclear 
multifragmentation.

In the next several years, planned experiments at FAIR (GSI) and at NICA (JINR) will 
strengthen the search for the CEP (and the first-order transition of the QCD 
phase diagram) by exploring regions of higher baryonic chemical potentials, and 
definitive conclusions concerning its possible existence and location are 
expected (a review on the experimental search of the CEP can be found in Ref.
\cite{Akiba:2015jwa}).

It is also important to point out that the location of the CEP is affected by 
several conditions like the isospin or strangeness content of the 
medium \cite{Costa:2013zca}, the role of the vector interaction in the medium 
\cite{Fukushima:2008wg,Costa:2015bza} or the presence of an external magnetic 
field \cite{Costa:2013zca,Costa:2015bza}. 
The possible location of the CEP will allow us to set stricter constraints on 
effective models. 

Considering all that has been mentioned above, together with its relevance for 
the understanding of the QCD phase diagram, in the present work we investigate 
the isentropic trajectories crossing the chiral phase transition around the CEP 
in both the crossover and first-order transition regions. 
We consider different scenarios of interest for the phase diagram obtained by
choosing different values of the isospin and the strangeness chemical potentials.
Finally, we explore the effects of the vector interaction and of an external 
magnetic field on the isentropic trajectories around the CEP.

\section{Model and Formalism}
\label{sec:model}

To investigate quark matter subject to strong magnetic fields, we will use the
PNJL model \cite{Fukushima:2003fw,Ratti:2005jh} with 2+1 flavors and a vector 
interaction. The PNJL Lagrangian in the presence of an external magnetic field 
is given by
\begin{align}
{\cal L} &= {\bar{\psi}} \left[i\gamma_\mu D^{\mu}-{\hat m}_f \right ] \psi
+ G_S \sum_{a=0}^8 \left [({\bar \psi} \lambda_ a \psi)^2 + 
	({\bar \psi} i\gamma_5 \lambda_a \psi)^2 \right ] \nonumber\\
&-G_D\left\{{\rm det} \left [{\bar \psi}(1+\gamma_5)\psi \right] + 
	{\rm det}\left [{\bar \psi}(1-\gamma_5)\psi\right] \right \} \nonumber\\
&- G_V \sum_{a=0}^8 \left [({\bar \psi}\gamma^{\mu}\lambda_a \psi)^2 + 
	({\bar \psi}\gamma^{\mu}\gamma_5\lambda_a\psi)^2 \right ] \nonumber\\
&+ \mathcal{U}\left(\Phi,\bar\Phi;T\right) - \frac{1}{4}F_{\mu \nu}F^{\mu \nu}
	\label{Pnjl}
\end{align}
The quark sector is described by the  SU(3) version of the NJL model where $G_S$ 
represents the four-Fermi coupling constant and $G_D$ denotes the ’t Hooft 
interaction strength \cite{Klevansky:1992qe,Hatsuda:1994pi,Buballa:2003qv}.

In Eq. (\ref{Pnjl}), $\psi = (u,d,s)^T$ is the quark field with three flavors, 
and ${\hat m}_f= {\rm diag}_f (m_u,m_d,m_s)$ is the corresponding (current) mass 
matrix.
$\lambda_0=\sqrt{2/3}I$, where $I$ is the unit matrix in the three-flavor space, 
and $0<\lambda_a\le 8$ denote the Gell-Mann matrices.

The coupling between the (electro)magnetic field $B$ and quarks, and that between the 
effective gluon field and quarks, is implemented  {\it via} the covariant 
derivative $D^{\mu}=\partial^\mu - i q_f A_{EM}^{\mu}-i A^\mu$, where $q_f$ is 
the quark electric charge ($q_d = q_s = -q_u/2 = -e/3$),  $A^{EM}_\mu$ and 
$F_{\mu \nu }=\partial_{\mu }A^{EM}_{\nu }-\partial _{\nu }A^{EM}_{\mu }$ 
are used to account for the external magnetic field, and 
$A^\mu(x) = g_{strong} {\cal A}^\mu_a(x)\frac{\lambda_a}{2}$, where
${\cal A}^\mu_a$ is the SU$_c(3)$ gauge field.
A static and constant magnetic field in the $z$ direction, 
$A^{EM}_\mu=\delta_{\mu 2} x_1 B$, is considered.
In the Polyakov gauge, 
$A^\mu = \delta^{\mu}_{0}A^0$,
with
$A^0 = - i A_4$. 
The trace of the Polyakov line defined by
$ \Phi = \frac 1 {N_c} {\langle\langle \mathcal{P}\exp i\int_{0}^{\beta}d\tau\,
A_4\left(\vec{x},\tau\right)\ \rangle\rangle}_\beta$
is the Polyakov loop which is the {\it exact} order parameter of the $\Z_3$ 
symmetric/broken phase transition in pure gauge.

To describe the pure gauge sector, we choose an effective potential, 
$\mathcal{U}\left(\Phi,\bar\Phi;T\right)$, which allows us to reproduce the results 
obtained in lattice calculations \cite{Roessner:2006xn}:
\begin{align}
	\frac{\mathcal{U}\left(\Phi,\bar\Phi;T\right)}{T^4}
	&=-\frac{a\left(T\right)}{2}\bar\Phi \Phi \nonumber\\
	&+b(T)\mbox{ln}\left[1-6\bar\Phi \Phi+4(\bar\Phi^3+ \Phi^3)-3(\bar\Phi \Phi)^2\right],\nonumber\\
	\label{Ueff}
\end{align}
where $a\left(T\right)=a_0+a_1\left(\frac{T_0}{T}\right)+a_2\left(\frac{T_0}{T}\right)^2$, 
$b(T)=b_3\left(\frac{T_0}{T}\right)^3$.
The choice of the parameters for the effective potential 
$\mathcal{U}\left(\Phi,\bar\Phi;T\right)$ is 
$a_0 = 3.51$, $a_1 = -2.47$, $a_2 = 15.2$, and $b_3 = -1.75$.
The parameter $T_0$ of the Polyakov potential defines the onset of deconfinement
and is fixed to $270$ MeV according to the critical temperature for the 
deconfinement in pure gauge lattice findings \cite{Kaczmarek:2002mc}. 
This potential was constructed in order to describe, for the pure 
Yang-Mills sector, the expectation value of the color traced Polyakov loop 
(which acts as an order parameter for the confinement-deconfinement transition). 
Other potentials were proposed in literature (see for example Ref.
\cite{Fukushima:2008wg}).
However, a new and consistent Polyakov-loop potential fitted to the latest 
lattice data that are continuum extrapolated and cover a large temperature range
is needed. This new effective potential will eventually change the behavior of 
the isentropic trajectories. Since this new potential is not yet available, we 
utilize the potential most commonly used in the literature given by 
Eq. (\ref{Ueff}).

The thermodynamical potential $\Omega$ is written as
\begin{align}
\Omega(T,\mu,B)&=2G_S \sum_{f=u,d,s}\ev{\bar{\psi}_f\psi_f}^2
-4G_D\ev{\bar{\psi}_u\psi_u}\nonumber\\
&\times\ev{\bar{\psi}_d\psi_d}\ev{\bar{\psi}_s\psi_s} 
-2G_V \sum_{f=u,d,s}\ev{\psi^{\dagger}_f\psi_f}^2\nonumber \\
&+\sum_{f=u,d,s}\pc{\Omega_{\text{vac}}^f+\Omega_{\text{med}}^f+\Omega_{\text{mag}}^f}
+{\cal U}(\Phi,\bar{\Phi},T)
\end{align}
where $\Omega_{\text{vac}}^f$, $\Omega_{\text{med}}^f$, and $\Omega_{\text{mag}}^f$
are the flavor contributions from the vacuum, medium and magnetic field, 
respectively, which can be found in the Appendix A.
The effective chemical potential, for each flavor, is given by
\begin{align}
\tilde{\mu}_f = \mu_f - 4G_V\rho_f,\,\,\,\, f = u,d,s.
\end{align}
The equation of state for the entropy density, $s$, is given by
\begin{align}
s(T,\mu,B) = -\frac{\partial\Omega}{\partial T}.
\end{align}

Since the model is not renormalizable, we regularize it by using a sharp cutoff 
in 3-momentum space, $\Lambda$, for the divergent ultraviolet integrals only. 
For our numerical calculations we adopt the parameter set obtained in Ref. 
\cite{Rehberg:1995kh}: $\Lambda= 602.3$ MeV, $G_S\Lambda^2= 1.835$, 
$G_D\Lambda^5 =12.36$, $m_u = m_d= 5.5$ MeV, $m_s= 140.7$ MeV.
In the vacuum ($T=\mu_q = 0$), this choice of parameters gives the quark masses 
$M^{vac}_q=M^{vac}_u = M^{vac}_d = 367.7$ MeV and $M^{vac}_s = 549.5$ MeV 
\cite{Rehberg:1995kh}.

\section{The isentropic trajectories}

We next analyze how the isentropic trajectories near the CEP are affected by the 
influence of a repulsive vector interaction and by the presence of an external 
magnetic field. We start with the case where the vector coupling is absent at zero 
magnetic field and use the results as benchmarks to discuss the effects of the 
vector interaction and of the presence of a magnetic field on the isentropes.
We will neglect the effect of the soft-mode fluctuations around the CEP. 
Indeed, soft modes associated with the CEP are important\footnote{
For the QCD CEP, the associated soft mode is a linear combination of 
fluctuations of the chiral condensate and the quark number density instead of 
pure chiral fluctuations \cite{Kamikado:2012cp}.};
however, their contribution is only important in a narrow region surrounding the 
CEP \cite{Fukushima:2009dx}.

\subsection{Results at $G_V = 0$}
\label{sec:Gv_zero}

\begin{figure*}[t!]
    \includegraphics[width=0.525\linewidth,angle=0]{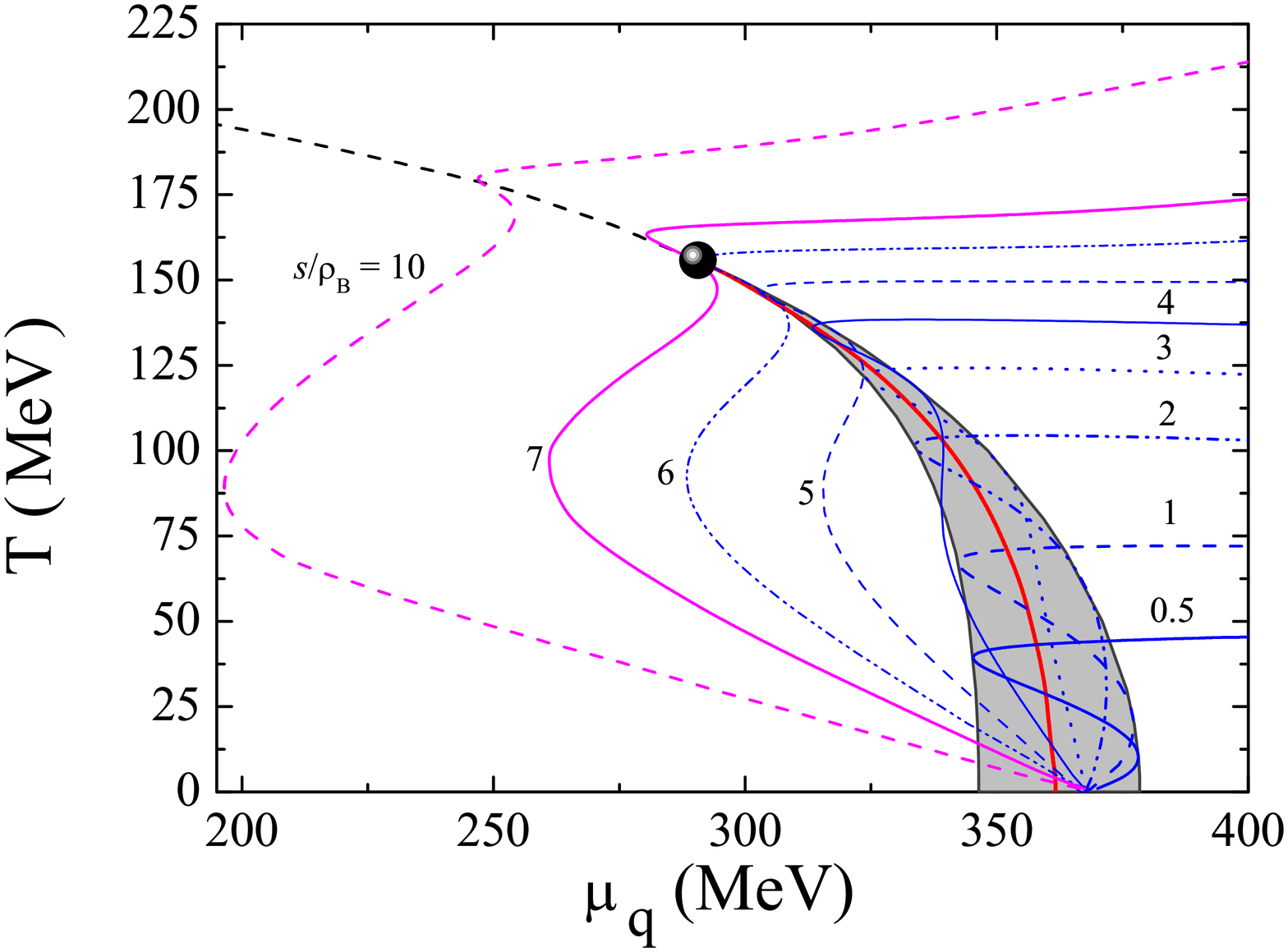}
    \hspace{-1cm}\includegraphics[width=0.525\linewidth,angle=0]{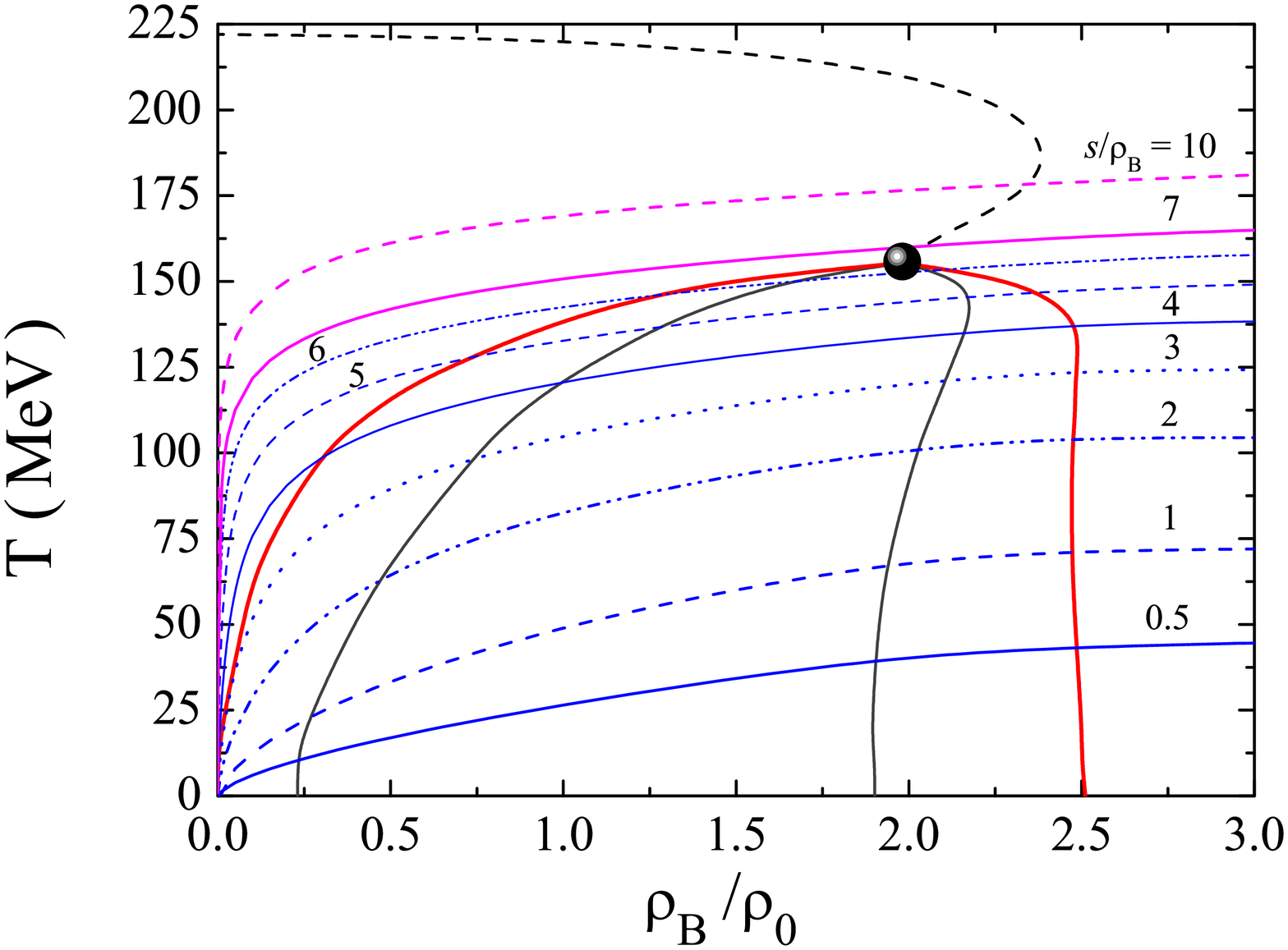}
    \caption{Left panel: QCD phase diagram in the $T-\mu_q$ plane.
    The full red line is the first-order chiral phase transition 
    and the gray region is the unstable region bounded by the spinodal lines
    $-$ spinodal region.
    Right panel: QCD phase diagram in the $T-\rho_B/\rho_0$ plane with 
    $\rho_0=0.16$ fm$^{-3}$.
    The full red line delineates the phase coexistence boundary (binodal line). 
    The metastable region is bounded externally by the the binodal line and is 
    separated from the unstable region by the spinodal lines (gray lines).
    For both panels the isentropic trajectories correspond to (from right) 
    $S/\rho_B=$ [0.5, 1, 2, 3, 4, 5, 6, 7, 10]. The isentropic trajectories in 
    magenta intersect the crossover line. 
    }
\label{fig:1}
\end{figure*}

In this section, we consider three different scenarios for the phase diagram:\\
\noindent 
\textit{Case} I $-$ equal quark chemical potentials ($\mu_u = \mu_d = \mu_s$). 
This scenario, used in most calculations, corresponds to zero charge (or isospin),
$\mu_Q = 0$, and zero strangeness chemical potential, $\mu_S = 0$. It also 
allows for isospin symmetry, $M_u = M_d$, and the net strange quark density, 
$\rho_s$, is nonzero.

\noindent 
\textit{Case} II$-$ equal $u$ and $d$ quark chemical potentials ($\mu_u=\mu_d$) 
and zero strange quark chemical potential ($\mu_s=0$). It corresponds to zero
charge (isospin) chemical potential, $\mu_Q=0$, and the strangeness chemical
potential is one third of the total baryonic chemical potential, 
$\mu_S=1/3\mu_B$. 
This is the relevant scenario to simulate matter created by ultra 
relativistic HIC. Indeed, since thermalization is reached within $10^{-22}$ sec 
(10-20 fm/c), the time scale of the strong interaction, the system is far from 
$\beta$-equilibrium. The net strange quark number should be zero before the 
beginning of hadronization in the expansion stage \cite{Greiner:1987tg} 
so, the strange quark density must be set to zero, $\rho_s=0$. 

\noindent
\textit{Case} III$-$ $\beta$-equilibrium matter corresponding to 
$\mu_u-\mu_d= \mu_{Q}=-\mu_e$ and $\mu_d=\mu_s$ ($\mu_S=0$). 

Next, we present our results for isentropic trajectories close to the CEP.
In Fig. \ref{fig:1} we plot the isentropic trajectories in the $T-\mu_q$ (left 
panel) and in the $T-\rho_B$ (right panel) planes for \textit{Case} I. 
The CEP is located at ($T^{\,CEP}=$ 155 MeV, $\mu_q^{CEP}=$ 291 MeV) 
(see Table \ref{table:CEP}). 
At zero temperature, the phase diagram presents a first-order phase transition 
with $\mu_q^{crit}=361$ MeV being the critical chemical potential where the 
transition takes place \cite{Buballa:2003qv,Costa:2010zw}.
At $\mu_q=0$, both chiral and deconfinement transitions are crossovers,
being the respective pseudocritical temperatures at $T_c^{\chi}=222$ MeV and 
$T^{\Phi}_c=210$ MeV \cite{Costa:2010zw}.

\begin{table}[b]
\begin{center}
  \begin{tabular}{|c||c|c|c|}
    \hline
    \hline
                        & $T$ [MeV] & $\mu_q$ [MeV] & $\rho_B/\rho_0$\\
    \hline
    \hline
    \textit{Case} I ($\mu_u=\mu_d=\mu_s$)   & 155 & 291 & $1.98$	\\
    \hline
    \textit{Case} II ($\rho_s=0$)           & 157 & 296 & $1.84$ \\
    \hline
    \textit{Case} III ($\beta$-equilibrium) & 146 & 308 & $1.71$ \\
    \hline
    \hline
  \end{tabular}
  \caption{Temperature, quark chemical potential and baryonic density for the CEP 
  in the different scenarios considered ($\rho_0=0.16$ fm$^{-3}$). 
  \label{table:CEP} }
\end{center}
\end{table}

By analyzing the behavior of the isentropic trajectories when $T\rightarrow0$, 
we note that all isentropic trajectories terminate at the same point of the 
horizontal axes: $T=0$ and $\mu_q=367.7$ MeV, with $\mu_q = M_q^{vac}>\mu_q^{crit}$. 
This combination ($T=0,\,\,\mu_q=367.7$ MeV) corresponds to the vacuum.
Indeed, for the chosen set of parameters, at $T = 0$, the first-order transition 
point satisfies the condition $\mu_q^{crit} < M^{vac}_q$ \cite{Buballa:2003qv}.
This scenario implies the existence of a strong first-order phase transition 
from the vacuum solution $M_q = M^{vac}_q$ into the partially chiral restored 
phase with $M_q$ small when compared with $M^{vac}_q$. At the transition point, 
the density jumps from zero to a relatively large value. 

According to the third law of thermodynamics, when $T\rightarrow0$, also 
$s\rightarrow0$, and therefore, for $s/\rho_B = const.$, we must require that 
$\rho_B\rightarrow0$, which is fulfilled when $\mu_q = M_q^{vac}$ 
\cite{Costa:2010zw}.
In fact, at $\mu_q^{crit}$ and $T=0$, the total baryonic density jumps from zero 
to $\sim 2.5\rho_0$, equally carried by quarks $u$ and $d$ 
(the density of strange quarks, $\rho_s$, is still zero, and we only have 
$\rho_s\ne0$ when $\mu_q > M_{s}$).

\begin{figure*}[t!]
    \includegraphics[width=0.525\linewidth,angle=0]{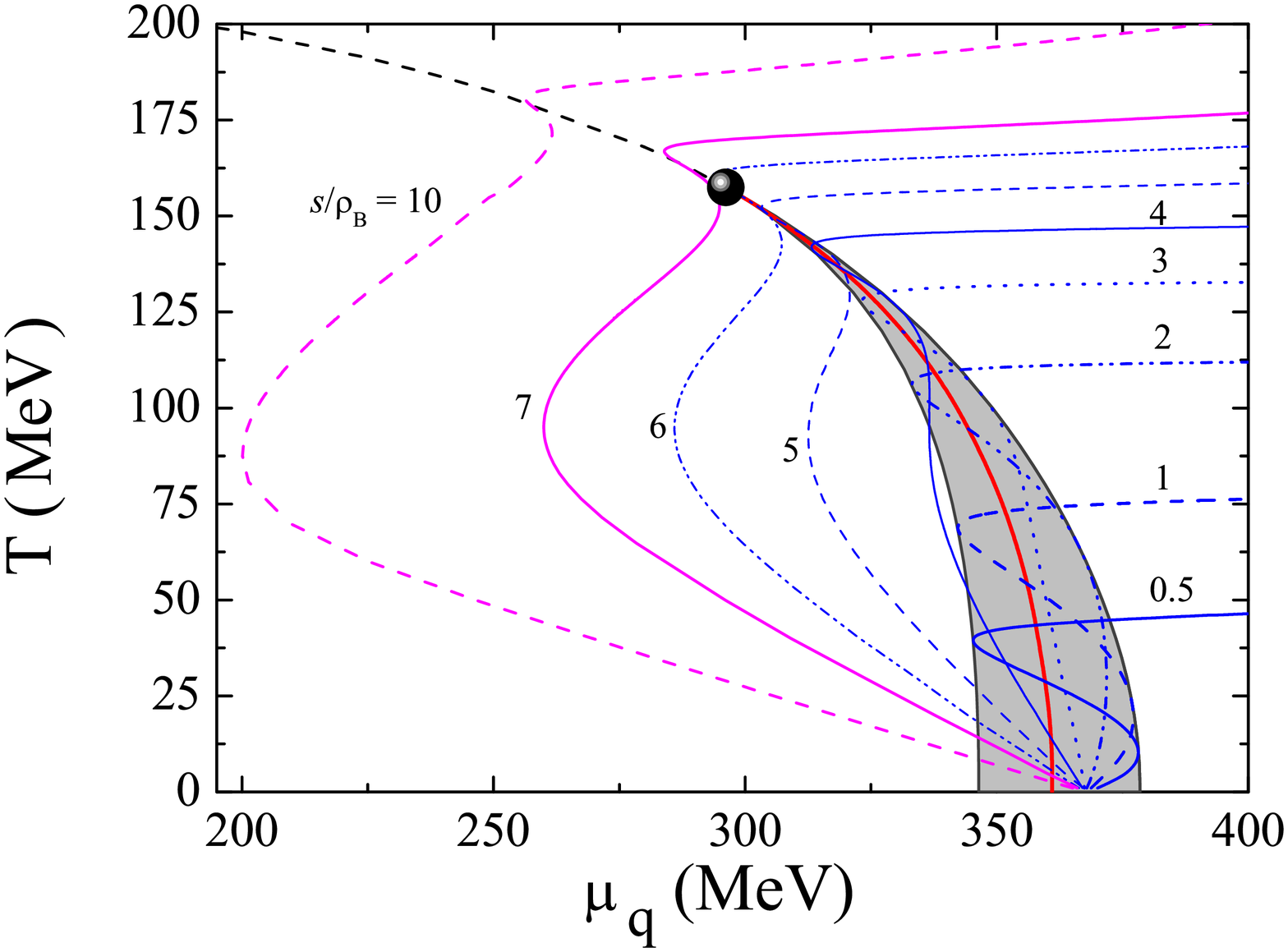}
    \hspace{-1cm}\includegraphics[width=0.525\linewidth,angle=0]{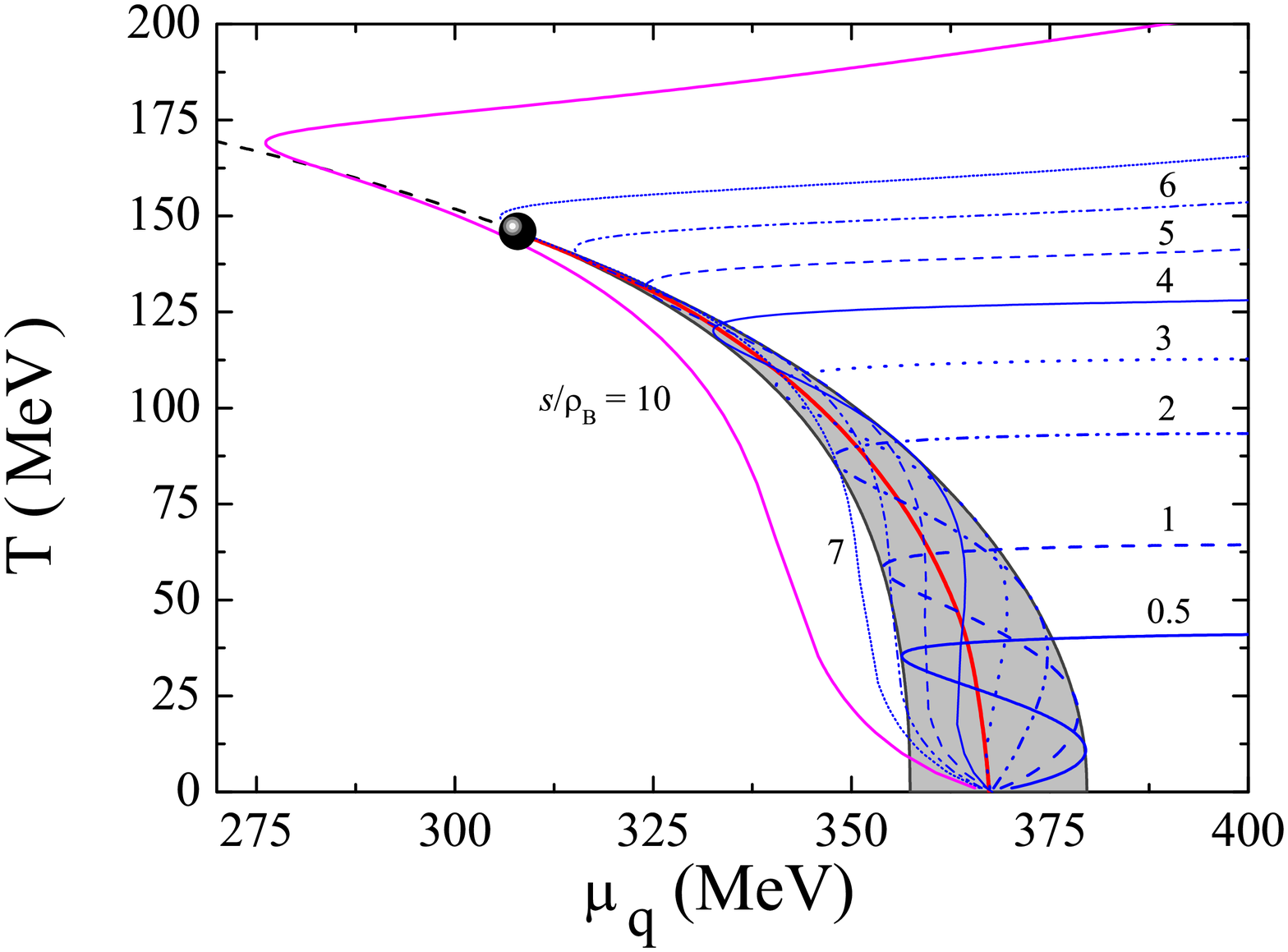}
    \caption{
		Left panel: Phase diagram in the $T-\mu_q$ plane for \textit{Case} II 
    ($\rho_s=0$). Right panel: Phase diagram in the $T-\mu_q$ plane for 
    \textit{Case} III ($\beta$-equilibrium). The isentropic trajectories in 
    magenta intersect the crossover line. 
    }
\label{fig:2}
\end{figure*}

In the vicinity of the first-order region, the isentropic trajectories with 
$s/\rho_B \lesssim 6$ come from the region of partially restored chiral
symmetry and reach the unstable region (spinodal region), bounded by the spinodal 
lines, going then along with it as $T$ decreases until it reaches $T = 0$
(see Fig. \ref{fig:1}, left panel). 
Taking the line for $s/\rho_B = 0.5$, it is seen that the isentropic trajectory 
intersects the spinodal line at ($T\approx44$ MeV, $\mu_q\approx372$ MeV) and
crosses the first-order line twice, at ($T\approx43$ MeV, $\mu_q\approx358$ MeV) 
and ($T\approx29$ MeV, $\mu_q\approx360$ MeV), as the temperature decreases in a 
``zigzag''-shaped trajectory bounded by the spinodal lines.
Now, taking the line with $s/\rho_B = 4$, it becomes interesting to note that the 
isentropic trajectory starts by having a behavior similar that of 
$s/\rho_B = 0.5$, but then it leaves the spinodal region. However, as the 
temperature decreases, the isentropic trajectory reaches the spinodal region 
again from lower values of $\mu_q$. This also happens for lines with 
$s/\rho_B = 5$ and $s/\rho_B = 6$.

For cases with $s/\rho_B > 6$, the isentropic trajectories, given by the curves 
in magenta, go directly through the crossover region (the crossover is defined 
as the zero of the 
$\partial^2{\left\langle \bar{q}_fq_f\right\rangle} /\partial T^2$, 
i.e., the inflection point of the light quark condensates 
$\left\langle {\bar{q}_fq_f}\right\rangle$, $f=u,\,d$), displaying a smooth 
behavior, and they reach the spinodal region from lower values of $\mu_q$. 
In the crossover region, the isentropic trajectories have a behavior
qualitatively similar  to that obtained in lattice calculations 
\cite{Borsanyi:2012cr,Ejiri:2005uv}. 

As already pointed out in Ref. \cite{Costa:2010zw}, no focusing of isentropic 
trajectories towards the CEP is seen but only smooth trajectories. 
The focusing effect was suggested in Ref. \cite{Nonaka:2004pg}, and it is argued 
that hot and dense QCD matter, as systems with a possible CEP of the same 
universality class as the three-dimensional Ising model, would exhibit 
focusing near the CEP: the CEP would act as an attractor for the isentropes
\cite{Nonaka:2004pg}.
However, several models do not exhibit the focusing effect near the CEP.
In Ref. \cite{Nakano:2009ps}, isentropic trajectories were investigated in a 
renormalization group approach applied to the quark-meson model, and no 
focusing was found. 
Indeed, while the critical behavior at the CEP is universal, the focusing 
effect is not, because the the entropy per baryon does not diverge at the CEP. 
The characteristic shape of the isentropic trajectories in the vicinity of the 
CEP can vary from model to model, even if they belong to the same 
universality class \cite{Nakano:2009ps}.

Next, we will proceed by investigating \textit{Case} II (see left panel of Fig.
\ref{fig:2}) due to its relevance heavy-ion collisions. 
As already pointed out in Ref. \cite{Fukushima:2009dx}, the phase diagram is only
slightly changed by the constraint $\rho_s=0$; in our case the CEP moves from 
($T^{\,CEP}= 155$ MeV, $\mu_q^{CEP}= 291$ MeV) to ($T^{\,CEP}= 157$ MeV, 
$\mu_q^{CEP}= 296$ MeV), as can be seen in Table \ref{table:CEP}.
In Ref. \cite{Fukushima:2009dx} it was also argued that the strangeness neutrality 
only has a perceptible effect on the isentropic trajectories at high temperatures. 
The reason for this behavior lies in the fact that the constituent mass of quark 
$s$, $M_s$, is still heavy around the chiral crossover when compared with the 
masses of the quarks $u$ and $d$, and consequently $\rho_s\approx 0$ even without 
a constraint, as long as $T$ is low and $\mu_q$ is smaller than $M_s$.

In right panel of Fig. \ref{fig:2}, we present the results concerning 
\textit{Case} III, matter in $\beta$-equilibrium, which has a large isospin 
asymmetry. The CEP for $\beta$-equilibrium occurs for a larger quark chemical 
potential, but also for a lower temperature when compared to the other scenarios.
The reason is related to the fact that matter in $\beta$-equilibrium being less 
symmetric, is less bound, and therefore, the transition to a chirally symmetric 
phase occurs at a lower temperature and density than in the symmetric case 
\cite{Costa:2013zca}.
On the other hand, at $T=0$, the spinodal region is reduced by 10 MeV when
compared with \textit{Cases} I and II, where the spinodal region occurs
between 346 and 378 MeV for both cases.
Concerning the isentropic trajectories, the behavior is qualitatively similar to
both cases previously studied, but they have some peculiarities, as can be seen 
in the right panel of Fig. \ref{fig:2}: 
i) all isentropic trajectories will end at $T=0$ and 
$\mu_q = M_q^{vac} = \mu_q^{crit} = 367.7$ MeV; and
ii) for the same values of $s/\rho_B$, the isentropic trajectories occur at 
lower temperatures, especially for lower values of $s/\rho_B$.

\subsection{Results at $G_V \ne 0$: The influence of the vector interaction 
on the isentropic trajectories}
\label{sec:Gv_dif_zero}

\begin{figure*}[t]
    \includegraphics[width=0.525\linewidth,angle=0]{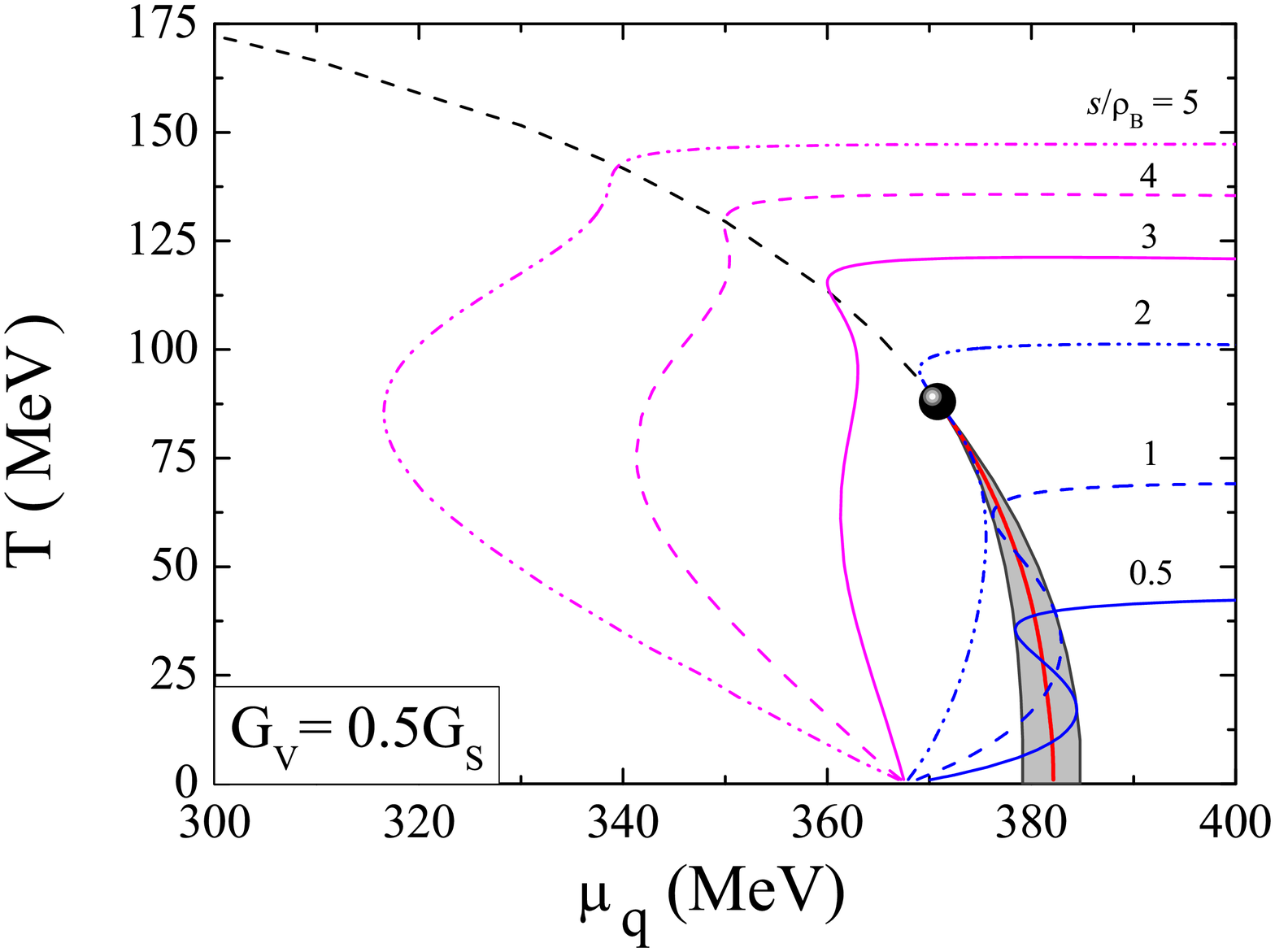}
    \hspace{-1cm}\includegraphics[width=0.525\linewidth,angle=0]{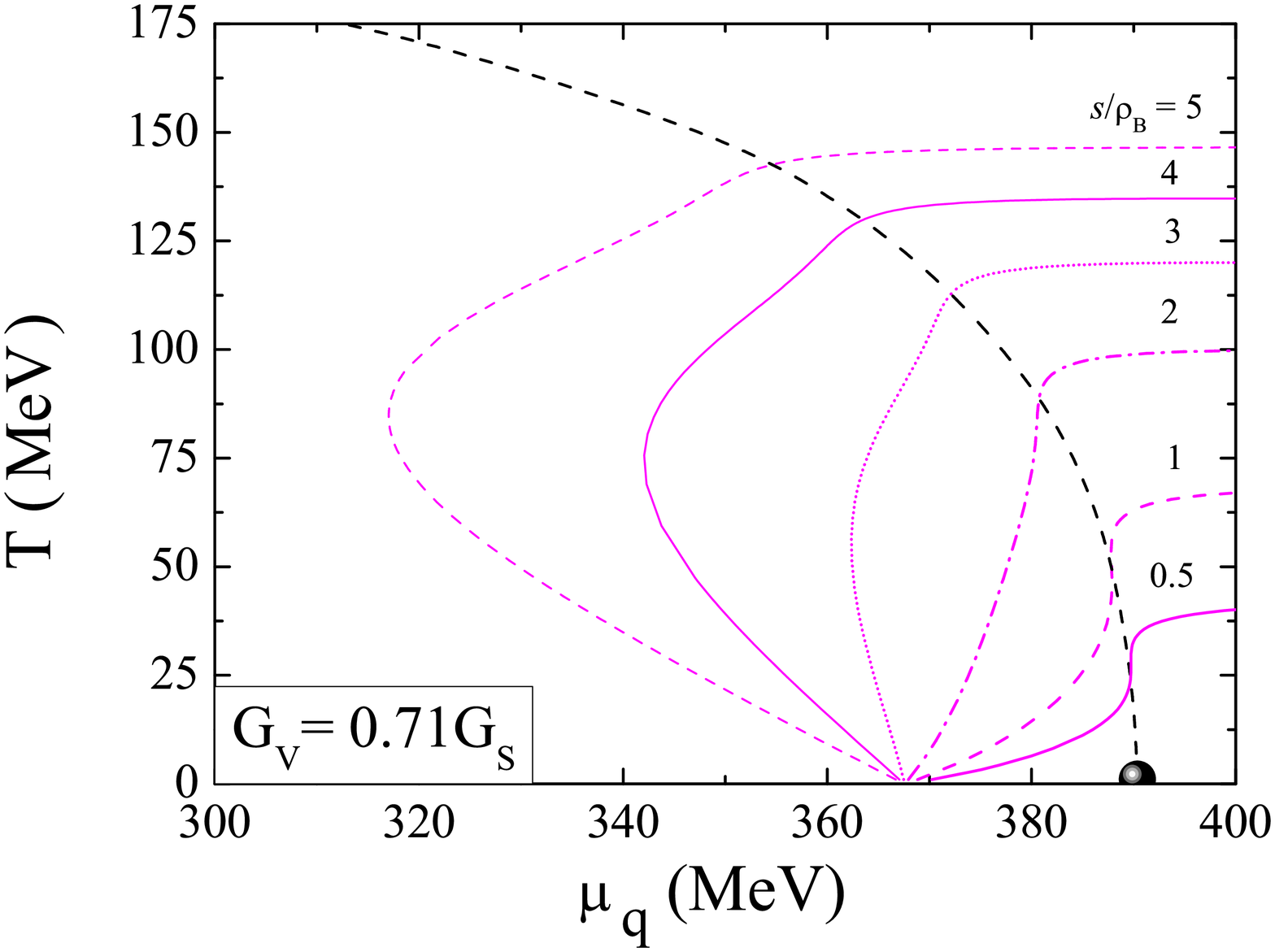}
    \caption{
    Phase diagram in the $T-\mu_q$ plane for $G_V=0.5G_S$ 
    (left panel) and $G_V=G_V^{crit}=0.71G_S$ (right panel). For 
    $G_V=G_V^{crit}=0.71G_S$ all the isentropic trajectories do not cross the 
    first-order transition region. 
    }
\label{fig:3}
\end{figure*}

In this section, we investigate the influence of the vector interaction in 
the isentropic trajectories. 
Henceforward, we restrict our study to \textit{Case} I ($\mu_u = \mu_d = \mu_s$).
The role of the vector interaction in the phase diagram was studied in detail 
in Ref. \cite{Fukushima:2008wg} and more recently in Ref. \cite{Costa:2015bza}, 
where the presence of an external magnetic field was also considered. 
It was shown that the CEP can be absent in the phase diagram when the value of 
the coupling $G_V$ is greater than the critical value of $G_V^{crit} 
\approx 0.71G_S$ within the present parametrization \cite{Costa:2015bza}.
Indeed, as the value of $G_V$ is increased from 0 to $G_V^{crit}$, the 
first-order phase transition is weakened: the CEP is located at lower 
temperatures and larger chemical potentials, but smaller densities. 
For $G_V=0.5G_S$, the CEP is located at 
($T^{\,CEP}= 88$ MeV, $\mu_q^{CEP}= 371$ MeV),
while at $G_V^{crit} \approx 0.71G_S$, it is located at 
($T^{\,CEP}\sim0$ MeV, $\mu_q^{CEP}= 390$ MeV) (see Table \ref{table:CEP_Gv}).

\begin{table}[b]
\begin{center}
  \begin{tabular}{|c||c|c|c|}
    \hline
    \hline
                  & $T$ [MeV] & $\mu_q$ [MeV] & $\rho_B/\rho_0$\\
    \hline
    \hline
    $G_V=0$ (\textit{Case} I)       & 155 & 291 & $1.98$	\\
    \hline
    $G_V=0.5G_S$  & 88 & 371 &	$\sim 1.4$\\
    \hline
    $G_V^{crit}=0.71G_S$ & $\sim0$ & 390 & $\sim 1.1$  \\
    \hline
    \hline
  \end{tabular}
  \caption{Temperature, quark chemical potential and baryonic density for the CEP 
  for different values of the vector coupling, $G_V$ ($\rho_0=0.16$ fm$^{-3}$). 
  \label{table:CEP_Gv} 
  }
\end{center}
\end{table}

In the left (right) panel of Fig. \ref{fig:3}, we present the results for the isentropic
trajectories when $G_V=0.5G_S$ ($G_V^{crit}\approx0.71G_S$). 
For $G_V=0.5G_S$ (left panel of Fig. \ref{fig:3}), their behavior is different 
from that obtained with $G_V=0$ (left panel of Fig. \ref{fig:1}): Due to the 
repulsive nature of the vector interaction, when $T=0$, the first-order phase 
transition only takes place at a critical potential $\mu_q^{crit} > M_q^{vac}$, 
meaning that the system will not form quark droplets; instead, the system has a 
homogeneous quark gas at low densities. 
The constituent mass of light quarks goes gradually down and the respective 
density starts to rise smoothly at $\mu_q\approx M_{q}^{vac}$, well before 
$\mu_q^{crit}$. 
Consequently, isentropic trajectories will not finish in the spinodal 
region as $T\rightarrow 0$ but for $\mu_q$ near $M_q^{vac}$. 
Taking the trajectory $s/\rho_B =1$ (dashed blue line), it intersects the 
first-order line three times, being the highest temperature crossing near the CEP. 
Then, as the temperature decreases in its path inside the spinodal region, it 
reaches the chemical potentials of both the lower and upper spinodal lines. 
At $T \approx 15$ MeV, the isentropic trajectory leaves the spinodal region to 
the chirally broken phase going to $\mu_q\sim M_q^{vac}$ when $T\rightarrow 0$.
The trajectory with $s/\rho_B =2$ goes through the crossover at $T\approx96$ MeV 
and $\mu_q\approx369$ MeV, very close to the CEP 
($T^{\,CEP}= 88$ MeV, $\mu_q^{CEP}= 371$ MeV), and comes into the spinodal 
region from lower values of $\mu_q$ just below the CEP. 
At $T\approx70$ MeV, the isentropic trajectory leaves the spinodal region,  
its behavior being similar to that of the line $s/\rho_B =1$.
The trajectories with $s/\rho_B \ge 3$ (in magenta) do not cross the spinodal 
region.

In the high-chemical-potential region, the isentropic trajectories behave similarly 
to \textit{Case} I (see also the left panel of Fig. \ref{fig:1}). In this region, 
the chiral symmetry is already restored for all scenarios, being the constituent
masses of the quarks close to their current values.  
This also happens when $G_V^{crit}\approx0.71G_S$ (see right panel of Fig. \ref{fig:3}), 
even if in this case the first-order phase transition no longer occurs and the 
transition to the partially chiral restored phase is a crossover, except at the CEP 
where the transition is of second-order.

\subsection{The influence of the magnetic field on the isentropic trajectories}
\label{sec:MF}

\begin{figure*}[t!]
    \includegraphics[width=0.525\linewidth,angle=0]{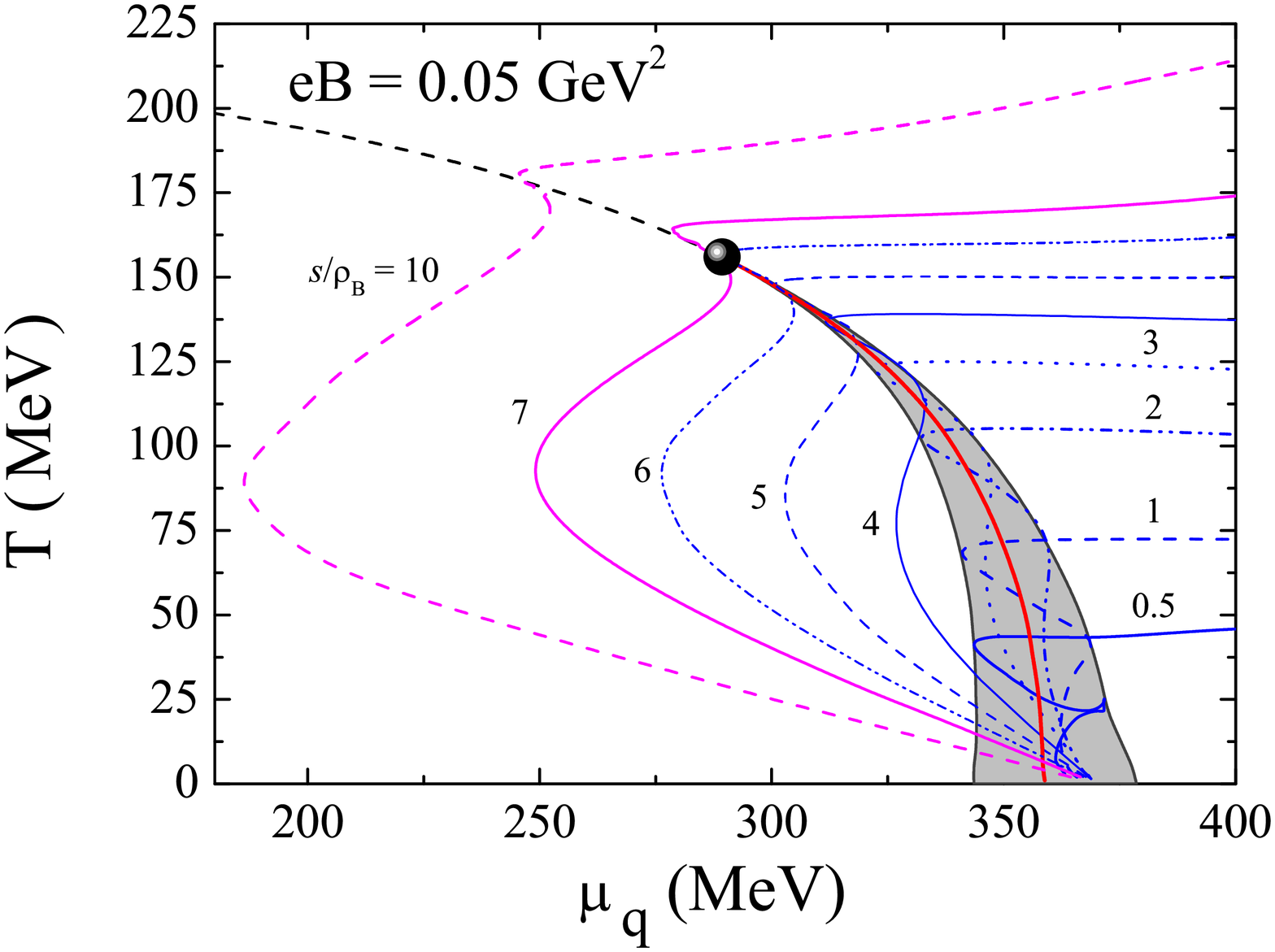}
    \hspace{-1cm}\includegraphics[width=0.525\linewidth,angle=0]{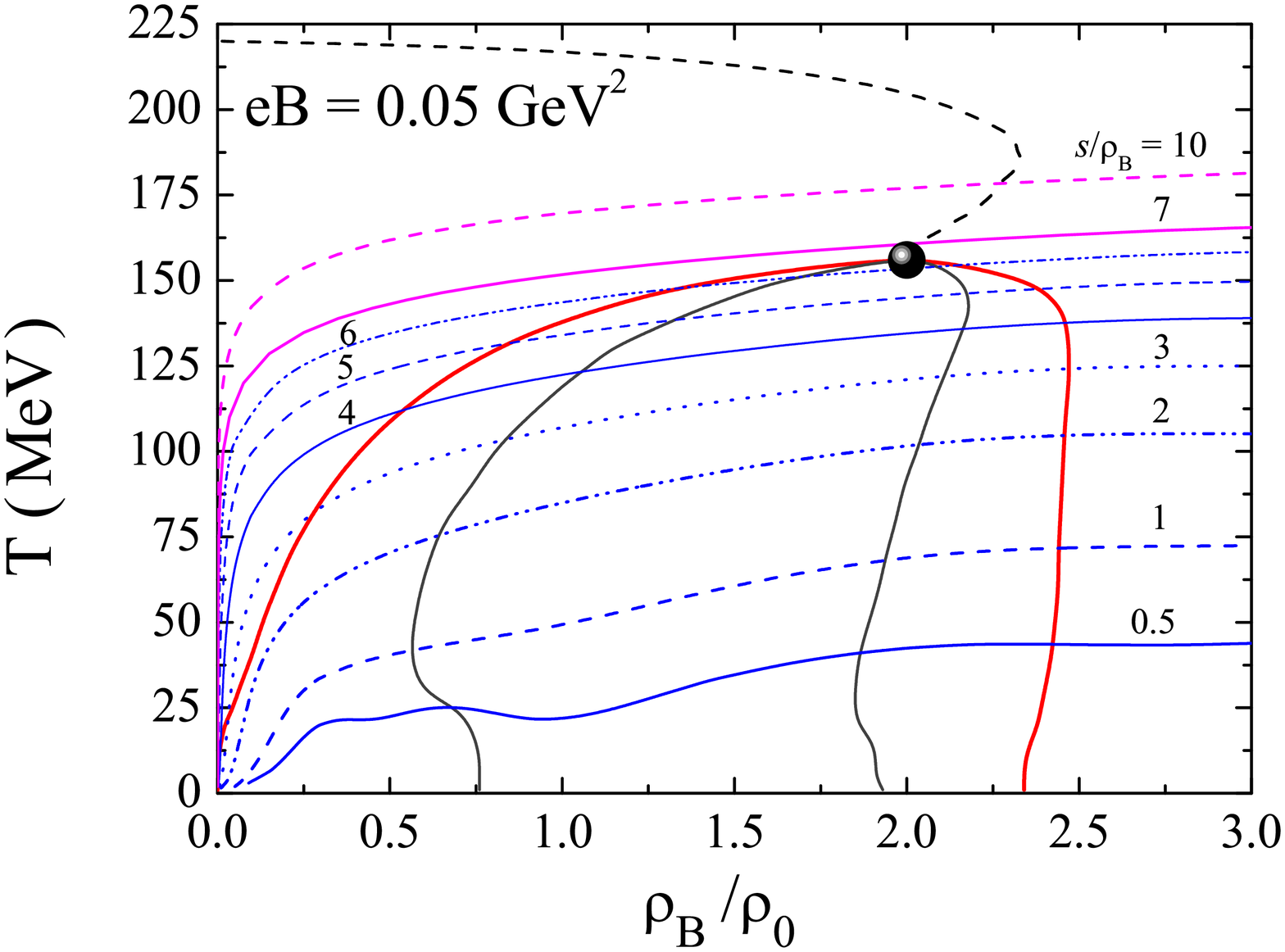}
    \caption{
    Left panel (right panel): Phase diagram in the $T-\mu_q$ ($T-\rho_B/\rho_0$) 
    plane for $eB = 0.05$ GeV$^{2}$.
    The isentropic trajectories correspond to (from right, counterclockwise) 
    $S/\rho_B=$ [0.5, 1, 2, 3, 4, 5, 6, 7, 10 ]. The isentropic trajectories in 
    magenta intersect the crossover line.  
    }
\label{fig:4}
\end{figure*}

The influence of a magnetic field gives rise to a magnetic catalysis (MC) 
effect$-$i.e., the enhancement of the quark condensate due to the magnetic field. 
Increasing the chemical potential and/or the temperature and increasing the magnetic 
field have competing effects: while high chemical potentials and/or temperatures 
favor the restoration of chiral symmetry, the magnetic field has an opposite 
effect \cite{Avancini:2011zz}.

The influence of external magnetic fields on the position of the CEP was 
investigated in detail for both NJL and PNJL models in Refs.
\cite{Avancini:2012ee,Costa:2013zca,Costa:2015bza}.
The main conclusions when $\mu_u=\mu_d=\mu_s$ were that the trend is similar for 
both models: as the intensity of the magnetic field increases until a certain 
value ($eB\sim0.4$ GeV$^2$ \cite{Costa:2013zca,Costa:2015bza}), the transition 
temperature increases and the baryonic chemical potential decreases.
For even stronger magnetic fields, when no inverse magnetic catalysis (IMC) 
effects are considered, both $T^{\,CEP}$ and $\mu_B^{CEP}$ increase 
\cite{Avancini:2012ee,Costa:2013zca}. 
This can be seen from Table \ref{table:CEP_eB}.
If the isospin symmetric matter scenario is taken, $\mu_u=\mu_d$ and $\mu_s= 0$, 
the influence of the magnetic field on the CEP is very similar to the previous 
one, the temperature being only slightly larger and the baryonic density only 
slightly smaller for the CEP's location \cite{Costa:2013zca}.

When IMC effects  \cite{Ferreira:2014kpa,Ferreira:2013tba} are taken into 
account, noticeable effects on the location of the CEP occur for sufficiently 
high values of the magnetic field: the CEP will now occur at increasingly 
smaller chemical potentials and at a practically unchanged temperature 
\cite{Costa:2015bza}. Under an increase of the magnetic field, the CEP eventually 
moves toward $\mu_B= 0$, and the deconfinement and chiral phase transitions 
should become of first-order as predicted by lattice calculations 
\cite{Endrodi:2015oba}.

\begin{table}[b]
\begin{center}
  \begin{tabular}{|c||c|c|c|}
    \hline
    \hline
                  & $T$ [MeV] & $\mu_q$ [MeV] & $\rho_B/\rho_0$\\
    \hline
    \hline
    $eB = 0$ (\textit{Case} I)       & 155 & 291 & $1.98$	\\
    \hline
    $eB = 0.05$ GeV$^{2}$ & 156 & 289 &	$\sim 2.0$\\
    \hline
    $eB = 0.3$  GeV$^{2}$ & 192 & 225 & $\sim 3.5$  \\
    \hline
    \hline
  \end{tabular}
  \caption{Temperature, quark chemical potential and baryonic density for the 
  CEP at $eB = 0$, $eB = 0.05$ GeV$^{2}$, and $eB = 0.3$ GeV$^{2}$ 
  ($\rho_0=0.16$ fm$^{-3}$). 
  \label{table:CEP_eB} 
  }
\end{center}
\end{table}

\begin{figure*}[tb]
    \includegraphics[width=0.525\linewidth,angle=0]{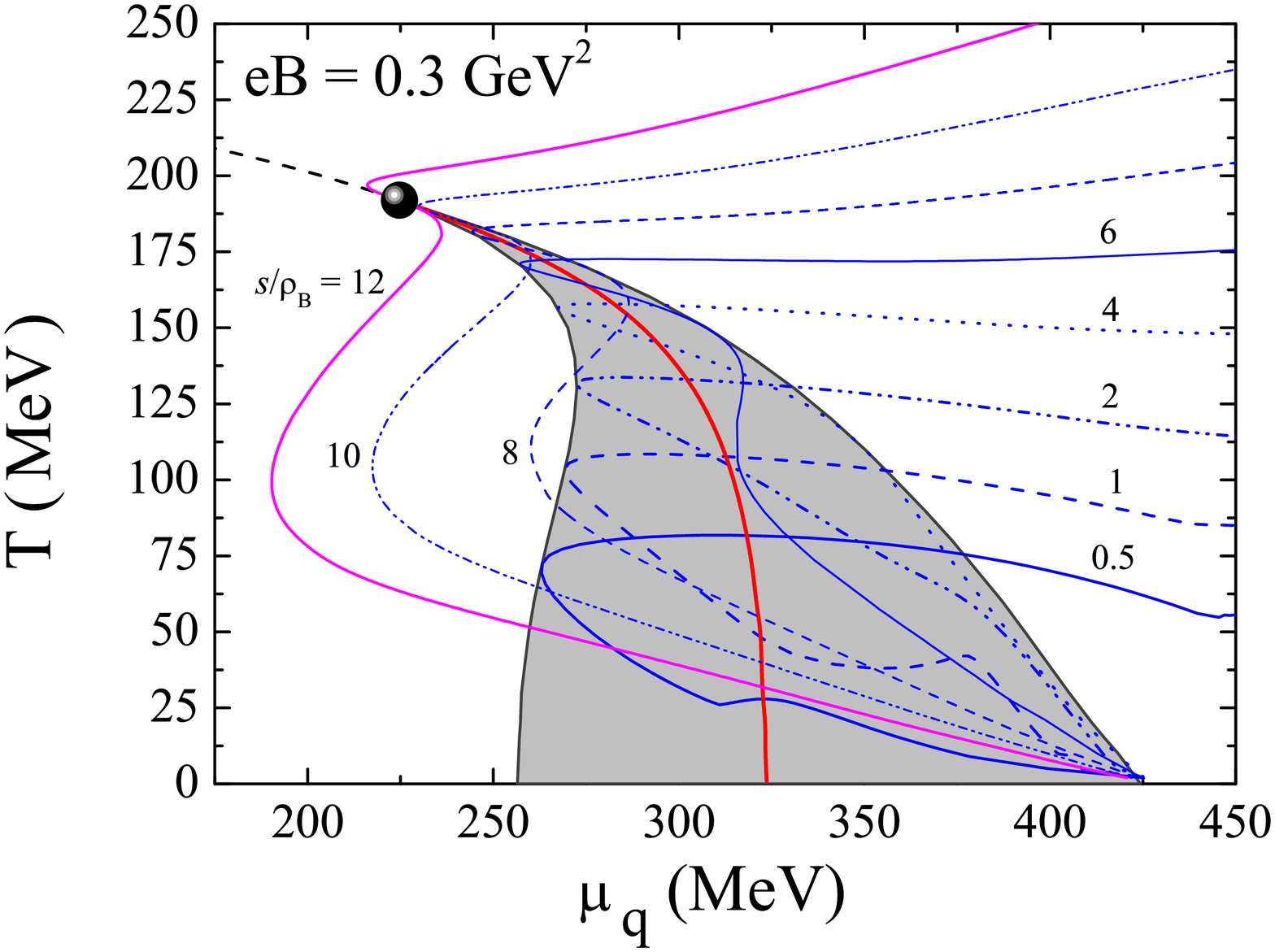}
    \hspace{-1cm}\includegraphics[width=0.525\linewidth,angle=0]{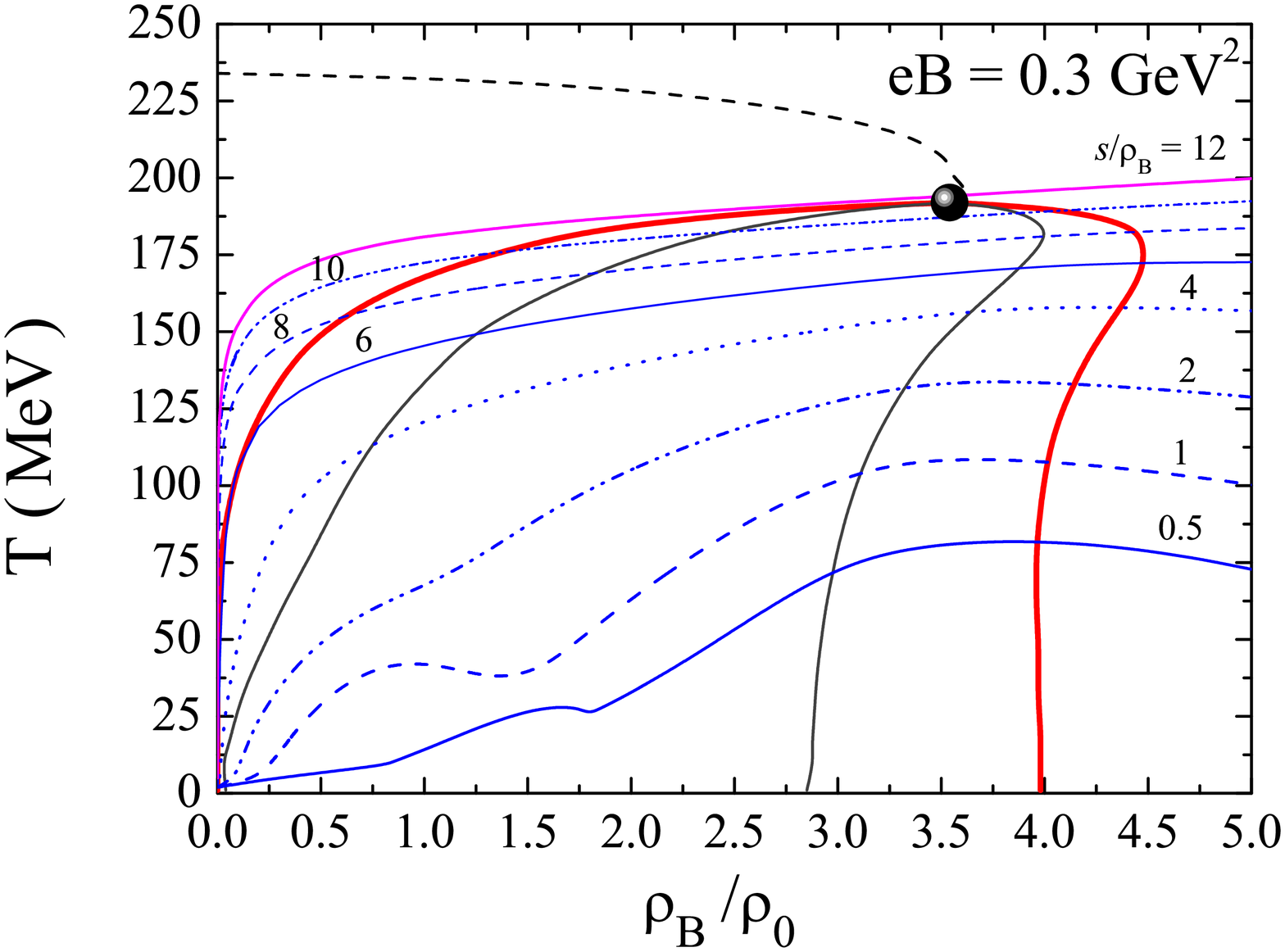}
    \caption{
    Left panel (right panel): Phase diagram in the $T-\mu_q$ ($T-\rho_B/\rho_0$) 
    plane for $eB = 0.3$ GeV$^{2}$.
    The isentropic trajectories correspond to (from right, counterclockwise) 
    $S/\rho_B=$ [0.5, 1, 2, 4, 6, 8, 10, 12]. The isentropic trajectories in 
    magenta intersect the crossover line.  
    }
\label{fig:5}
\end{figure*}

To investigate the influence of the magnetic field on the behavior of isentropic 
trajectories we take $eB = 0.05$ GeV$^2$, a relatively small magnetic field, 
and $eB = 0.3$ GeV$^2$ (a magnetic field that could occur at LHC experiments,
even if the magnetic field can only be produced for a short period of time in HIC), 
which are plotted in Figs. \ref{fig:4} and \ref{fig:5}, respectively. 
No IMC effects on the location of CEP will be considered.
From the panels of both figures, we conclude that as the magnetic field 
increases, the spinodal region is enlarged, especially for $eB=0.3$ GeV$^{2}$: 
the extension of the spinodal region at $T=0$ goes from $\Delta\mu_q=32$ MeV 
for $eB=0$ to $\Delta\mu_q= 167$ MeV for $eB=0.3$ GeV$^{2}$.
A complete study of the effect of the magnetic field intensities on the EOS at 
$T=0$ was performed in Refs. \cite{Menezes:2014aka,Costa:2015bza}. 
As pointed out, for some ranges of the magnetic field, several first-order phase 
transitions occur which disappear when the temperature is slightly increased, so 
they will not influence the behavior of the chosen isentropic trajectories. 
In fact, like in our case for $eB=0.05$ GeV$^{2}$, the filling of the Landau 
levels originates Haas–van Alphen oscillations (the effect of Landau 
quantization of charged particles' energy). Larger values of $B$ increase the 
amplitude of the fluctuations and reduce their number, because fewer Landau 
levels are involved \cite{Menezes:2014aka}. 
The consequence is that the larger the intensity of the magnetic field, the 
greater the difficulty in restoring chiral symmetry.
On the other hand, by increasing the temperature, thermal energy can become of 
the order of the level spacing and washes out these Landau quantization effects. 
Nevertheless, at sufficiently low temperatures, Landau quantization still 
manifests itself by the appearance of oscillations in physical quantities like 
the pressure.

The investigated isentropic trajectories are quite affected by the growth of the 
spinodal region as would be expected: there is an entrainment of the 
isentropes to higher temperatures within this region as it grows, due to the 
increasing of the magnetic field, as can be seen by comparing the left panels of 
Figs. \ref{fig:4} and \ref{fig:5}. 
Together with the shift of the CEP (see Table \ref{table:CEP_eB}) to higher 
(lower) temperatures (chemical potentials) by increasing magnetic fields 
(making the spinodal region bigger), also the isentropic trajectories are pushed 
to higher temperatures in the transition region, especially when 
$eB = 0.3$ GeV$^{2}$.

However, outside of the spinodal region and for high chemical potentials, the 
isentropes have almost the same temperature as those with $eB=0$, in particular 
for $eB = 0.05$ GeV$^{2}$. 
This can be understood by considering the fact that at high temperatures and 
chemical potentials, the restoration of chiral symmetry already took place and 
the magnetic catalysis effect is almost suppressed: 
due to the increasing temperature, more Landau Levels will be filled, when 
compared with low temperatures, being the system less affected by the magnetic field.
For lower temperatures isentropes with $s/\rho_B < 4$, at $eB=0.3$ GeV$^{2}$, are 
more affected by the magnetic field (as can be seen in left panel of Fig. \ref{fig:5}), 
where they can decrease as $\mu_q$ is increased.

As in the previous sections, no focusing of isentropic trajectories towards the 
CEP is seen with increased magnetic field but only smooth trajectories, 
as with the results obtained in the model SU (2). 

Finally, from the right panels of Figs. \ref{fig:4} and \ref{fig:5}, it is seen 
that the baryonic density area of the transition region increases as the 
magnetic field increases. This is the result of the strengthening of the 
first-order transition due to the magnetic field. As already mentioned, 
stronger magnetic fields induce larger spacing between the Landau levels, and 
consequently, it is more difficult to restore chiral symmetry.
On the other hand, for high magnetic fields, the lower-density spinodal line 
is pulled to very small densities, and some isentropes (see the case 
for $s/\rho_B=4$ in the right panel of Fig. \ref{fig:5}) can reach the spinodal 
region for small values of the density. 
With this behavior, it is expected that the enhancement of fluctuations of 
observables caused by spinodal instabilities, like strangeness, will be more 
easily detected in the presence of strong magnetic fields.


\section{Conclusions}

In the present work we have investigated the isentropic trajectories around the 
CEP. 
The isentropic trajectories are very interesting because the hydrodynamical 
expansion of a HIC fireball nearly follows trajectories of constant entropy.
New insights about the QCD phase diagram can be obtained by investigating these 
possible paths for the hydrodynamic evolution of a thermal medium created in the
collisions.
An unambiguous experimental identification of a first-order phase transition
(if it exists) from the hadronic phase to the deconfined phase is one important 
goal of HIC experiments at present and future facilities. 
However, the first-order transition can be affected by the isospin or strangeness 
content of the medium, by the role of the vector interaction in the medium, or 
by external conditions, like the presence of an external magnetic field.

We considered different scenarios of interest for the phase diagram by choosing 
matter with different content of strangeness and isospin (the $\beta$-equilibrium
case having the largest isospin asymmetry).
We also explored the effects of the vector interaction and of an external 
magnetic field on the isentropic trajectories around the CEP.

More asymmetric matter weakens the first-order transition: the CEP for
matter in $\beta$-equilibrium occurs at lower temperatures and densities than 
the symmetric cases ($\mu_u=\mu_d=\mu_s$ and $\mu_u=\mu_d$, $\mu_s=0$); 
at $T=0$ the spinodal region is also smaller for matter in $\beta$-equilibrium.
Concerning the isentropic trajectories, the behavior is qualitatively similar 
for all cases investigated, but for the same values of $s/\rho_B$, the isentropic 
trajectories occur at lower temperatures for matter in $\beta$-equilibrium.

When a vector interaction is introduced, the first-order transition becomes
even weaker, which is reflected in the CEP location (it can even disappear 
in the $\mu_q$ axis) and in an increasingly smaller area of instability as 
$G_V$ grows. 
In the high chemical potential region the isentropic trajectories behave 
similarly to the case without vector interaction once the chiral symmetry is 
already restored for all scenarios and the influence of vector interaction 
fades out.  

On the other hand, the influence of a strong magnetic field on the phase diagram
is reflected in the strengthening of the first-order transition with the 
respective enlargement of the spinodal region and the displacement of the CEP 
to higher temperatures and lower chemical potentials.  
The isentropic trajectories are entrained for higher temperatures, and the ones 
with lower values of $s/\rho_B$ at lower temperatures are more affected by the 
magnetic field, especially for higher values of $B$.
For high magnetic fields, the lower-density spinodal line is pulled to small 
densities, and some isentropes can reach the spinodal region for small values of 
the density. It is then expected that there will be an enhancement of 
fluctuations of observables caused by spinodal instabilities in the presence of 
strong magnetic fields at lower densities.

\vspace{1cm}

\begin{center}
{\bf Acknowledgment: } 
\end{center}

I would like to thank C. Provid\^{e}ncia and J. Moreira for helpful discussions.
This work was supported by ``Fundação para a Ciência e Tecnologia", Portugal, 
under Grant No. SFRH/BPD/102273/2014.


\appendix
\section{}
\label{Appendix} 

The flavor contributions from vacuum $\Omega_{\text{vac}}^f$, medium 
$\Omega_{\text{med}}^f$, and magnetic field $\Omega_{\text{mag}}^f$ 
\cite{Menezes:2008qt,Menezes:2009uc} are given by
\begin{align}
 \Omega_{\text{vac}}^f&=-6\int_{\Lambda}\frac{d^3p_f}{(2\pi)^3}E_f,
\end{align}
\begin{align}
\Omega_{\text{med}}^f&=-T\frac{|q_fB|}{2\pi}\sum_{n=0}\alpha_n\int_{-\infty}^{+\infty}\frac{dp_z^i}{2\pi}\pc{Z_{\Phi}^+(E_f)+Z_{\Phi}^-(E_f)},
\end{align}
\begin{align}
\Omega_{\text{mag}}^f&=-\frac{3(|q_f|B)^2}{2\pi^2}\pr{\zeta^{'}(-1, x_f)-\frac{1}{2}(x_f^2-x_f)\ln x_f+\frac{x_f^2}{4}},
\end{align}
where $E_f=\sqrt{(p_z^i)^2+M_f^2+2|q_f|Bk}$ , $\alpha_0=1$ and $\alpha_{k>0}=2$,
$x_f=M_f^2/(2|q_f|B)$, and $\zeta^{'}(-1, x_f)=d\zeta(z, x_f)/dz|_{z=-1}$, 
where $\zeta(z, x_f)$ is the Riemann-Hurwitz zeta function. 
$Z_{\Phi}^+$ and $Z_{\Phi}^-$ read
\begin{align}
 Z_{\Phi}^+=\ln\Big{\{} 1+3\bar{\Phi}e^{-\beta (E_f-\mu_f)}&+3\Phi e^{-2\beta (E_f-\mu_f)}\nonumber \\
&+e^{-3\beta (E_f-\mu_f)}\Big{\}},
\end{align}
\begin{align}
Z_{\Phi}^-=\ln\Big{\{} 1+3\Phi e^{-\beta (E_f+\mu_f)}&+3\bar{\Phi} e^{-2\beta (E_f+\mu_f)}\nonumber \\
&+e^{-3\beta (E_f+\mu_f)}\Big{\}}.
\end{align}
The quark condensates $\ev{\bar{\psi}_f\psi_f}$ are given by 
$
\ev{\bar{\psi}_f\psi_f}=\ev{\bar{\psi}_f\psi_f}_{\text{vac}}+\ev{\bar{\psi}_f\psi_f}_{\text{mag}}+\ev{\bar{\psi}_f\psi_f}_{\text{med}},
$
where
\begin{align}
\ev{\bar{\psi}_f\psi_f}_{\text{vac}}&=-6\int_{\Lambda}\frac{d^3p}{(2\pi)^3}\frac{M_f}{E_f},
\end{align}
\begin{align}\ev{\bar{\psi}_f\psi_f}_{mag}=-\frac{3m_f|q_f|B}{2\pi^2}&\Big{[} 
  \ln\Gamma(x_i)-\frac{1}{2}\ln(2\pi)+x_i\nonumber\\
  &-\frac{1}{2}(2x_f-1)\ln(x_f)\Big{]},
\end{align}
\begin{align}
\ev{\bar{\psi}_f\psi_f}_{\text{med}}&=\frac{3(|q_f|B)^2}{2\pi}\nonumber\\
&\times\sum_n\alpha_n\int_{-\infty}^{+\infty}
\frac{dp_z^i}{2\pi}\pc{
f_{\Phi}^+(E_f)+f_{\Phi}^-(E_f)}.
\end{align}
The distribution functions $f_\Phi^{+}$ and $f_\Phi^{-}$ are
{\small
\begin{align}
 f_{\Phi}^+(E_f)&=\frac{\bar{\Phi} e^{-\beta (E_f-\mu_f)}+2\Phi e^{-2\beta (E_f-\mu_f)}+e^{-3\beta 
 (E_f-\mu_f)}  
 }
 {1+3\bar{\Phi} e^{-\beta (E_f-\mu_f)}+3\Phi e^{-2\beta (E_f-\mu_f)}+e^{-3\beta (E_f-\mu_f)}},\nonumber\\
\end{align}
}
{\small
\begin{align}
f_{\Phi}^-(E_f)&=\frac{\Phi e^{-\beta (E_f+\mu_f)}+2\bar{\Phi} e^{-2\beta (E_f+\mu_f)}+e^{-3\beta (E_f+\mu_f)}}
 {1+3\Phi e^{-\beta (E_f+\mu_f)}+3\bar{\Phi}e^{-2\beta (E_f+\mu_f)}+e^{-3\beta (E_f+\mu_f)}}.\nonumber\\
\end{align}
}
By employing a mean-field approach the effective quark masses, $\Phi$, and $\bar{\Phi}$ can be obtained self-consistently from
\begin{align}
M_f=m_f-4G_S\ev{\bar{\psi}_f\psi_f}-2G_D\epsilon_{ijk}\ev{\bar{\psi}_j\psi_j}\ev{\bar{\psi}_k\psi_k}\nonumber\\
\end{align}
and
\begin{align}
\frac{\partial{\cal U}}{\partial \Phi}=\frac{\partial{\cal U}}{\partial \bar{\Phi}}=0.
\end{align}


\end{document}